\documentclass[prd,aps,floats,twocolumn]{revtex4}

\usepackage{graphicx}
\usepackage{amssymb} 
\usepackage{amsmath}

\begin{document}

\title{Pulsar timing arrays as imaging gravitational wave telescopes:
  \\ angular resolution and source (de)confusion}

\author{Latham Boyle$^{1}$ and Ue-Li Pen$^{2}$}

\affiliation{$^1$Perimeter Institute for Theoretical Physics,
  Waterloo, Ontario, Canada \\
  $^2$Canadian Institute for Theoretical Astrophysics, Toronto,
  Ontario, Canada}

\date{May 2012}

\begin{abstract}
  Pulsar timing arrays (PTAs) will be sensitive to a finite number of
  gravitational wave (GW) ``point'' sources (e.g. supermassive black
  hole binaries).  $N$ quiet pulsars with accurately known distances
  $d_{pulsar}$ can characterize up to $2N/7$ distant chirping sources
  per frequency bin $\Delta f_{gw}=1/T$, and localize them with
  ``diffraction limited'' precision $\delta\theta\gtrsim(1/{\rm SNR})
  (\lambda_{gw}/d_{pulsar})$.  Even if the pulsar distances are {\it
    poorly} known, a PTA with $F$ GW frequency bins can still
  characterize up to $(2N/7)(1-\frac{1}{2F})$ sources per bin, and the
  quasi-singular pattern of timing residuals in the vicinity of a GW
  source still allows the source to be localized quasi-topologically
  within roughly the smallest quadrilateral of quiet pulsars that
  encircles it on the sky, down to a limiting resolution
  $\delta\theta\gtrsim(1/{\rm SNR})\sqrt{\lambda_{gw}/d_{pulsar}}$.
  PTAs may be unconfused, even at the lowest GW frequencies: in that
  case, standard analysis techniques designed to detect a stochastic
  GW background would be incomplete and suboptimal, whereas matched
  filtering could provide more information and sensitivity.
\end{abstract}                        
\maketitle 

\section{Introduction}
\label{Introduction}

Our Local Group of galaxies is sprinkled with milli-second pulsars --
natural clocks of extraordinary stability.  Gravitational waves (GWs)
passing through the Milky Way, after being generated {\it e.g.}\ by
the inspiral of two supermassive black holes in a distant galaxy,
generate fluctuations in the time of arrival (TOA) of the pulses at
the Earth \cite{Sazhin, Detweiler, hellings83, Demorest09, Verbiest}.
In the future, we are likely to detect such GWs via their coherent
imprint on the TOA fluctuations from a collection of pulsars
distributed on the sky: a ``pulsar timing array'' (PTA).  Several
collaborations are actively monitoring pulsars for this purpose,
including NANOGrav, EPTA, PPTA and IPTA (see \cite{NANOGrav,
  EPTAstochastic, PPTAstochastic, PPTAptsource, IPTA, Demorest09} for
a review).  Much research has focused on using PTAs to study
stochastic GW backgrounds (\cite{FosterBacker,jenet06, vanh09} and
references therein), and upper bounds are improving\cite{demorest12}.
Recently, various authors have begun to study the ability of PTAs to
detect and characterize individual GW point sources \cite{jenet04,SVV,
  SesanaVecchio, FinnLommen, Lommen2011,
  PPTAptsource,DengFinn,CorbinCornish,Lee}.

Continuing in this direction, this paper is concerned with
conceptually clarifying the theoretical behavior and capabilities of
PTAs as GW point source telescopes.  We address two related issues.
(i) A PTA may be sensitive to so many GW sources that it becomes
``confused'' -- {\it i.e.}\ unable to disentangle and individually
characterize the sources.  When does a PTA become ``confusion
limited'' rather than sensitivity limited?  How many GW sources is it
capable of individually characterizing? \cite{SesanaVecchio} (ii) When
a set of GW point sources can be individually characterized, how well
can their angular positions be determined? \cite{jenet04,SVV,
  SesanaVecchio, FinnLommen, Lommen2011,
  PPTAptsource,DengFinn,CorbinCornish,Lee}

Regarding issue (i) we will see that PTAs with many pulsars can
characterize many GW sources per GW frequency bin; the traditional
rule of thumb that a GW detector becomes confused when there is more
than about one GW source per GW frequency bin is too pessimistic for
PTAs.  Regarding issue (ii) we must distinguish pulsars whose
distances are known {\it accurately} or {\it poorly} relative to
$\lambda_{gw}/(1-{\rm cos}\,\theta)$, where $\lambda_{gw}$ is the
gravitational wavelength and $\theta$ is the angle between pulsar and
source.  Pulsars with accurately known distances can angularly
localize a GW source very precisely; each such pulsar acts like a
single baseline of a diffraction-limited radio interferometer array --
with the radio wavelength replaced by the gravitational wavelength,
and the length of the radio baseline replaced by the distance from the
pulsar to the Earth!  The contribution from pulsars with poorly known
distances is more interesting: due to a quasi-singularity in the
pattern of timing residuals near the location of the GW source, the
source can still be localized surprisingly well, for reasons that have
less to do with diffraction, and more to do with topology (also see
\cite{FinnLommen}).

The paper is organized as follows.  Section \ref{BasicFormalism}
establishes our notation and basic formalism; this section is mainly a
rederivation of previously known results, but casts them in a compact
and general form that will be important for our analysis in the
following sections.  Our new results are contained in Sections
\ref{AccuratelyKnownDistances} and \ref{PoorlyKnownDistances}, in
which we obtain and interpret useful and new analytic expressions for
both the angular resolution and confusion limits of a PTA, when the
distances to the pulsars are accurately or poorly known, respectively.
Finally, Section \ref{Discussion} discusses some of the key
implications of our results, and highlights a number of open questions
for future research.

\section{Basic Formalism}
\label{BasicFormalism}

We label the 3 spatial directions with the latin indices
$\{i,j,k,l,m=1,2,3\}$, raised and lowered with $\delta^{ij}$ and
$\delta_{ij}$.  The $N$ pulsars in the network are labelled by the
greek indices $\{\alpha,\beta=1,\ldots,N\}$, raised and lowered with
$\delta^{\alpha\beta}$ and $\delta_{\alpha\beta}$.  We go to the
trouble of introducing raised and lowered indices here simply to make
use of the convenient Einstein summation convention: repeated indices
(one upper, one lower) are summed.

A gravitational wave on Minkowski space is described in
transverse-traceless (TT) gauge \cite{MTW} by the line element
$ds^{2}=-dt^{2}+[\delta_{ij}+2h_{ij}]dx^{i}dx^{j}$.  In this gauge,
the $\vec{x}={\rm constant}$ worldlines are timelike geodesics; along
such worldlines, the proper time $\tau$ is the coordinate time $t$.
To avoid notational clutter, let us start with just a single
gravitational plane wave travelling in the $\hat{n}$ direction:
\begin{equation}
  \label{hij}
  h_{ij}(t,\vec{x})=\int_{-\infty}^{\infty}df
  \tilde{h}_{ij}(f){\rm e}^{2\pi if(\hat{n}\cdot\vec{x}-t)};
\end{equation}
it is straightforward to extend the following analysis to a sum of
$m=1,\ldots,M$ plane waves, each travelling in a different direction
$\hat{n}_{m}$; this extension is discussed below.  Throughout this
paper, we use ``dot product'' notation to mean contraction with the
unperturbed 3-metric $\delta_{ij}$: $\vec{a}\cdot\vec{b}\equiv
\delta_{ij}a^{i}b^{j}$; and hats denote unit 3-vectors:
$\hat{a}\cdot\hat{a}=1$.  To avoid confusion, please note: in this
paper, since $\hat{n}$ is the direction of gravitational wave
propagation, the direction to the gravitational wave source is
$-\hat{n}$, {\it not} $+\hat{n}$.

If an electromagnetic flash is emitted from position $\vec{x}_{i}$ at
time $t_{i}$, what is its arrival time $t$ at position $\vec{x}_{f}$?
If we define $\vec{x}_{fi}\equiv
\vec{x}_{f}-\vec{x}_{i}=x_{fi}\hat{x}_{fi}$ then, at {\it zeroth}
order ({\it i.e.}\ in the absence of gravitational waves) the answer
is $t_{0}=t_{i}+x_{fi}$.  Solving the geodesic equation to first order
in $h_{ij}$ yields the perturbed result $t=t_{0}+\delta t$ where:
\begin{eqnarray}
  \label{t_first_order}
  \delta t=\!\!\int\!df
  \frac{i\tilde{h}_{ij}(f)\hat{x}_{\!fi}^{i}\hat{x}_{\!fi}^{j}
  [{\rm e}^{2\pi if(\hat{n}\cdot\vec{x}_{f}-t_{0})}
  \!-\!{\rm e}^{2\pi if(\hat{n}\cdot\vec{x}_{i}-t_{i})}]}
  {2\pi f(1\!-\!\hat{n}\cdot\hat{x}_{\!fi})}.
\end{eqnarray}
Now consider an observer at fixed spatial position $\vec{x}=\vec{0}$
receiving signals from $\alpha=1,\ldots,N$ pulsars at spatial
positions $\vec{r}_{\alpha}=r_{\alpha}\hat{r}_{\alpha}$.  For pulsar
$\alpha$, the TOA fluctuation $\delta t_{\alpha}(t_{0})$, as a
function of the unperturbed TOA $t_{0}$, is
\begin{equation}
  \label{deltat(t)}
  \delta t_{\alpha}(t_{0})=\int_{-\infty}^{\infty}df
  \delta\tilde{t}_{\alpha}(f){\rm e}^{-2\pi if t_{0}}
\end{equation}
where $\delta\tilde{t}_{\alpha}(f)$, the Fourier transform of the TOA
fluctuation, is given by
\begin{equation}
  \label{deltat(f)}
  \delta \tilde{t}_{\alpha}(f)=
  \frac{i\tilde{h}_{ij}(f)\hat{r}_{\alpha}^{i}\hat{r}_{\alpha}^{j}
    [1-{\cal P}_{\alpha}(f)]}{2\pi f(1\!+\!\hat{n}\!\cdot\!\hat{r}_{\alpha})}
\end{equation}
and, for later convenience, we have defined the phase
\begin{equation}
  \label{P_alpha}
  {\cal P}_{\alpha}(f)\equiv
  {\rm e}^{2\pi if r_{\alpha}(1+\hat{n}\cdot\hat{r}_{\alpha})}.
\end{equation}


The {\it measured} TOA fluctuations $s_{\alpha}(t_{0})$ from pulsar
$\alpha$ are gravitational wave signal $\delta t_{\alpha}(t_{0})$ plus
noise $n_{\alpha}(t_{0})$:
\begin{equation}
  s_{\alpha}(t_{0})=\delta t_{\alpha}(t_{0})+n_{\alpha}(t_{0}).
\end{equation}
(As a caveat, some of the TOA fluctuation ``signal'' may not be
attributed to gravitational waves, but instead must be absorbed into
determining the parameters of the pulsar timing model which accounts
{\it e.g.}\ for the relative motion of the pulsars and the Earth.
This caveat is important for GWs with periods of 1 year, or $>10$
years, but may be ignored otherwise -- particularly for the more
conceptual questions that are the focus of this paper.)  We take the
noise to be stationary and gaussian, so it is characterized by its
correlation function $C_{\alpha\beta}(T)$ or, equivalently, its
spectral density $S_{\alpha\beta}(f) =\tilde{C}_{\alpha\beta}(f)$:
\begin{subequations}
  \begin{eqnarray}
    C_{\alpha\beta}(T)&=&\overline{n_{\alpha}(t_{0}+T)n_{\beta}(t_{0})}
    \\
    \delta(f-f')S_{\alpha\beta}(f)&=&
    \overline{\tilde{n}_{\alpha}^{\ast}(f)\tilde{n}_{\beta}(f')}.
  \end{eqnarray}
\end{subequations}
Then, given any two functions $g^{(1)}(t)$ and $g^{(2)}(t)$, we
can define their natural noise-weighted inner product to be
\begin{equation}
  (g^{(1)}|g^{(2)})=\int_{-\infty}^{\infty}df\,\tilde{g}_{\alpha}^{(1)}(f)^{\ast}
  [S^{-1}(f)]^{\alpha\beta}\tilde{g}_{\beta}^{(2)}(f).
\end{equation}
We will assume that the noise is approximately uncorrelated between
different pulsars: $S_{\alpha\beta}(f)=S_{\alpha}(f)
\delta_{\alpha\beta}$.  (This approximation is a common one in the
pulsar literature.  At the moment, it is justified by the fact that
the terrestrial time standards used for pulsar timing are accurate to
about 10 ns -- {\it i.e.}\ the terrestrial clock error is small
relative to the timing fluctuations in the quietest current pulsars,
and relative to the gravitational-wave induced fluctuations currently
being sought.  Furthermore, it may be that future technological
improvements keep the terrestrial time standards perpetually ahead of
the accuracy needed for gravitational wave detection; but if this ever
fails to be the case, one would have to determine whether the noise
between different pulsars is significantly correlated, and incorporate
those correlations into the analysis that follows.)  Under the
assumption that the noise is uncorrelated between different pulsars,
matched filtering will detect a given gravitational wave signal with
expected signal-to-noise ratio squared (SNR$^{2}$) given by
\begin{subequations}
  \label{SNRsqr}
  \begin{eqnarray} 
    {\rm SNR}^{2}\!&\!=\!&\!(\delta t|\delta t)
    =\sum_{\alpha=1}^{N}{\rm SNR}_{\alpha}^{2} \\
    \label{SNRsqr_alpha}
    {\rm SNR}_{\alpha}^{2}\!&\!=\!&\!\int_{-\infty}^{\infty}df
    \frac{|\delta\tilde{t}_{\alpha}(f)|^{2}}{S_{\alpha}(f)}
  \end{eqnarray}
\end{subequations}
where $\delta\tilde{t}_{\alpha}(f)$ is given by (\ref{deltat(f)}).
When a gravitational wave signal (which depends on various parameters
$\xi^{k}$) is detected with sufficient SNR, the likelihood function
({\it i.e.}\ the probability of the observed data stream as a function of
the underlying signal parameters $\xi^{k}$) may be approximated as a gaussian
$\propto{\rm exp}[-(1/2)\xi^{k}\Gamma_{kl}\xi^{l}]$ near its peak, and
the expected inverse covariance matrix is the Fisher information
matrix, given by
\begin{equation}
    \Gamma_{kl}=\Big(\frac{\partial t}{\partial \xi^{k}}\Big|
    \frac{\partial t}{\partial \xi^{l}}\Big).
\end{equation}
The Fisher information matrix quantifies the accuracy with which the
parameters $\xi^{k}$ may be inferred from the observed signal.  For an
introduction to signal analysis and Fisher matrices, and their use in
gravitational wave detection, see Refs.~\cite{WainsteinZubakov, Finn,
  CutlerFlanagan}.  We are interested, in particular, in the angular
resolution of a PTA.  Define an orthonormal triad from $\hat{n}$ and
two other unit vectors $\hat{m}_{\bar{\mu}}$ ($\bar{\mu}=1,2$); let
$\gamma^{\,\bar{\mu}}$ be the rotation angle around
$\hat{m}_{\bar{\mu}}$.  The $2\times2$ angular part of $\Gamma_{kl}$
is
\begin{subequations}
  \begin{eqnarray}
    \label{Fisher}
    \Gamma_{\bar{\mu}\bar{\nu}}\!&\!=\!&\!
    \left(\frac{\partial[\delta t]}{\partial\gamma^{\,\bar{\mu}}}\Big|
      \frac{\partial[\delta t]}{\partial\gamma^{\,\bar{\nu}}}\right)
    =\sum_{\alpha=1}^{N}\Gamma_{\bar{\mu}\bar{\nu}}^{\alpha} \\
    \label{Fisher_alpha}
    \Gamma_{\bar{\mu}\bar{\nu}}^{\alpha}\!&\!=\!&\!
    \int_{-\infty}^{\infty}\!df
    \frac{\partial[\delta\tilde{t}_{\alpha}(f)]}
    {\partial\gamma^{\,\bar{\mu}}}^{\ast}\!\!\frac{1}{S_{\alpha}(f)}
    \frac{\partial[\delta\tilde{t}_{\alpha}(f)]}{\partial\gamma^{\,\bar{\nu}}}.
  \end{eqnarray}
\end{subequations}
To evaluate these angular derivatives, we act with the infinitessimal
rotation $R_{ij}\approx\delta_{ij}-\epsilon_{ijk}
\hat{m}_{\bar{\mu}}^{k}\gamma^{\,\bar{\mu}}$ on the gravitational wave
field, but {\it not} on the pulsar positions: {\it e.g.}\
$\partial(1+\hat{n}\cdot\hat{r}_{\alpha})/\partial\gamma^{\,\bar{\mu}}
=\epsilon_{ijk}\hat{n}^{i}\hat{r}_{\alpha}^{j}\hat{m}_{\bar{\mu}}^{k}$
and
$\partial[\tilde{h}_{ij}(f)\hat{r}_{\alpha}^{i}\hat{r}_{\alpha}^{j}]/
\partial\gamma^{\,\bar{\mu}}=2\tilde{h}_{il}(f)
\epsilon_{jk}^{\;\;\;\,l}\hat{r}_{\alpha}^{i}\hat{r}_{\alpha}^{j}\hat{m}_{\bar{\mu}}^{k}$.
In this way, we find
\begin{equation}
  \label{deltatDeriv}
  \frac{\partial[\delta\tilde{t}_{\alpha}(f)]}{\partial\gamma^{\,\bar{\mu}}}
  =\frac{i [{\cal A}(f)+{\cal B}(f)]}{2\pi f}
\end{equation}
where ${\cal A}$ comes from differentiating the phase ${\cal
  P}_{\alpha}$ in (\ref{deltat(f)}), and ${\cal B}$ comes from
differentiating everything else:
\begin{subequations}
  \label{AB}
  \begin{eqnarray}
    \label{A}
    {\cal A}\!&\!\equiv\!&\!
    2\pi if r_{\alpha}\frac{\tilde{h}_{ij}\hat{r}_{\alpha}^{i}\hat{r}_{\alpha}^{j}
      \hat{r}_{\alpha}^{k}\hat{n}^{l}\hat{m}_{\bar{\mu}}^{m}\epsilon_{klm}}
    {(1+\hat{n}\cdot\hat{r}_{\alpha})}{\cal P}_{\alpha} \\    
    \label{B}
    {\cal B}\!&\!\equiv\!&\!
    \frac{\tilde{h}_{ij}\hat{r}_{\alpha}^{i}\hat{r}_{\alpha}^{k}\hat{m}_{\bar{\mu}}^{l}}
    {1\!+\!\hat{n}\cdot\hat{r}_{\alpha}}
    \Big[2\epsilon^{j}_{\;\,kl}
    \!-\!\frac{\hat{r}_{\alpha}^{j}\hat{n}^{m}\epsilon_{klm}}
    {1\!+\!\hat{n}\cdot\hat{r}_{\alpha}}\Big]
    [1\!-\!{\cal P}_{\alpha}].\qquad
  \end{eqnarray}
\end{subequations}
In understanding the meaning of these equations, we should distinguish
two cases: (i) pulsars whose distances $r_{\alpha}$ are known {\it
  accurately} relative to $\lambda_{gw}/
(1+\hat{n}\cdot\hat{r}_{\alpha})$, so ${\cal P}_{\alpha}$ is known;
and (ii) pulsars whose distances $r_{\alpha}$ are known {\it poorly}
relative to $\lambda_{gw}/(1+\hat{n}\cdot\hat{r}_{\alpha})$, so ${\cal
  P}_{\alpha}$ is essentially a random phase.  We consider these two
cases in turn.

\section{Pulsars whose distances are accurately known}  
\label{AccuratelyKnownDistances}

First consider a monochromatic gravitational plane wave of frequency
$f_{gw}=c/\lambda_{gw}$, and a pulsar whose distance $r_{\alpha}$ is
known accurately relative to $\lambda_{gw}/(1+\hat{n}
\cdot\hat{r}_{\alpha})$.  Then, if $2\pi r_{\alpha}/\lambda_{gw}\gg1$,
the ${\cal A}$ term dominates the ${\cal B}$ term in
Eq.~(\ref{deltatDeriv}).  Combining Eqs.~(\ref{deltat(f)},
\ref{SNRsqr}--\ref{AB}), we obtain
\begin{equation}
  \label{diffraction_limited}
  \frac{\Gamma_{\bar{\mu}\bar{\nu}}^{\alpha}}{{\rm SNR}_{\alpha}^{2}}
  \approx\big(\frac{2\pi r_{\alpha}}{\lambda_{gw}}\big)^{2}
  \frac{(\hat{r}_{\alpha}^{i}\hat{n}^{j}\hat{m}_{\bar{\mu}}^{k}\epsilon_{ijk})
    (\hat{r}_{\alpha}^{i'}\hat{n}^{j'}\hat{m}_{\bar{\nu}}^{k'}\epsilon_{i'j'k'})}
  {|1-{\cal P}_{\alpha}(f_{gw})|^{2}}.
\end{equation}
Since the second fraction on the right-hand side of this equation is
typically ${\cal O}(1)$, this says that when a pulsar at a well known
distance $r_{\alpha}$ registers a gravitational wave with
signal-to-noise level SNR$_{\alpha}$, its contribution to
$\Gamma_{\bar{\mu}\bar{\nu}}^{\alpha}$ is typically
$\Gamma^{\alpha}_{\bar{\mu}\bar{\nu}} \sim(2\pi
r_{\alpha}/\lambda_{gw})^{2}{\rm SNR}_{\alpha}^{2}$.  In other words,
each such pulsar acts just like one of the baselines of a radio
interferometer array; but, in this analogy, the radio waves are
replaced by gravitational waves, and the baselines are of galactic
length scales and extend in all three spatial dimensions -- a
remarkable instrument!

Now consider multiple GW sources.  At the low GW frequencies probed by
PTAs (where the expected GW point sources are supermassive black hole
binaries, far from final merger) the frequency of each gravitational
plane wave drifts negligibly over the observation timescale
$T\sim10~{\rm yrs}$; and over the light travel time from the pulsars
to the Earth, the frequency drift or ``chirp'' is approximately
linear: $h_{ij}(t,\vec{x})={\rm Re} \{\hat{h}_{ij}{\rm e}^{-2\pi
  i\chi(\tau)}\}$, where $\chi(\tau)\equiv f_{0}\tau
+\frac{1}{2}\dot{f}\tau^{2}$, $\tau\equiv t-\hat{n} \cdot\vec{x}$, and
$\{\hat{h}_{ij}, f_{0}, \dot{f}\}$ are constants (see \cite{jenet04,
  SesanaVecchio, EPTAstochastic} for more discussion of the drift
specifics).  The induced timing residuals for pulsar $\alpha$ are a
sum of two peaks in frequency space: an ``Earth term'' at frequency
$f_{0}$, and a ``pulsar term'' at frequency $f_{0}-\dot{f}r_{\alpha}
(1+\hat{n}\cdot\hat{r}_{\alpha})$; if $\dot{f}$ is large enough ({\it
  i.e.}\ for supermassive black hole binaries of sufficiently high
mass, sufficiently close to merger) these two peaks may lie in
separate frequency bins \cite{jenet04}.  The number of such GW sources
that may be individually characterized by a PTA may be determined via
the following counting argument.  First recall that a PTA that
collects data for a total timespan $T$ ({\it e.g.}\ 10 years) cannot
distinguish two GW frequencies separated by less than $\Delta f\sim
1/T$; thus we should think of the PTA's GW frequency spectrum as being
divided into bins of finite width ($\sim\Delta f$).  To fully specify
the pattern of timing residuals, we must provide the following
information: in every GW frequency bin, and for each ``Earth term'' in
that bin, we give the associated propagation direction $\hat{n}$,
frequency derivative $\dot{f}$, and two complex amplitudes ({\it
  i.e.}\ an the amplitude and phase for both polarization modes), for
a total of 7 real numbers.  On the other hand, since the angular
dependence of Eq.~(\ref{deltat(f)}) contains spherical harmonics of
arbitrarily high angular momentum order, the number of independent
measurements collected by the PTA is simply $2N$ per GW frequency bin
-- namely, the measured amplitude and phase of the timing residuals,
for each pulsar, in each GW frequency bin \footnote{If
  Eq.~(\ref{deltat(f)}) only contained spherical harmonics up to
  angular momentum order $l\leq l_{max}$, the number of independent
  measurements collected by the PTA would be limited by
  $(l_{max}+1)^{2}$, the total number of spherical harmonics up to
  this order.}.  To completely characterize the individual sources,
the independent measurements must outnumber the parameters to be
determined; that is, the PTA can characterize up to an average of
$2N/7$ chirping GW point sources per GW frequency bin.  For
simplicity, this argument neglects ``boundary effects'' coming from GW
sources for which the ``Earth term'' lies within the detectable
frequency range, while the ``pulsar term'' does not, or vice versa.
If we assume that all of the GW sources are monochromatic
($\dot{f}=0$), the maximum number that can be characterized improves
only slightly to $2N/6$ per GW frequency bin, but the fitting
procedure becomes much easier since we can treat each GW frequency bin
independently.  If the PTA can disentangle and characterize the
individual sources, one expects the angular resolution to be
diffraction limited $\delta\theta\gtrsim(1/{\rm SNR})
(\lambda_{gw}/r_{pulsar})$ [the more precise expectation is given by
Eq.~(\ref{diffraction_limited})].

Typically, to be in the ``accurately measured'' category, a pulsar's
distance would have to be known to better than a gravitational
wavelength ({\it e.g.}\ a few parsecs). Although most pulsar distances
are poorly known, often only to a factor of 2 using dispersion
measure, the quiet pulsars most relevant to GW timing sometimes allow
more accurate distance measurements via ``timing parallax'' or
ordinary imagining parallax. Some of the most impressive pulsar
distances that have been obtained to date are $156.3\pm1.3~{\rm pc}$
(to J0437-4715) \cite{Deller2008}, $262\pm5~{\rm pc}$ (to B0950+08),
$433\pm8~{\rm
  pc}$ (to B0809+74) \cite{Brisken2002}, $361\pm9~{\rm pc}$ (to
B1929+10) \cite{Chatterjee2004}, and $950\pm20~{\rm pc}$ (to
(B2045-16) \cite{Chatterjee2009}. These high accuracies are currently
only available for rather nearby pulsars; but note the following two
points. First of all, remember that the relevant accuracy threshold
for the pulsar distance $r_{\alpha}$ is not actually $\lambda_{gw}$,
but rather $\lambda_{gw}/(1+\hat{n} \cdot\hat{r}_{\alpha})$, which is
$\gg\lambda_{gw}$ for pulsars that happen to be located in nearly the
same direction as the gravitational wave source. As a result, the
distances to such pulsars can be of benefit, even if they are known to
an accuracy considerably worse than the relevant gravitational
wavelength. Second, looking toward the future, there is still the
potential for significantly improved pulsar distance measurements. On
one hand, an instrument like the Square Kilometer Array could
dramatically improve both the accuracy and distance reach of distance
pulsar measurements via both timing and imaging parallax: see {\it
  e.g.}\ \cite{Smits2011}.
On the other hand, there is promising potential for interstellar
holographic distance measurements: low frequency VLBI combined with
cyclic spectroscopy \cite{2011MNRAS.416.2821D} may allow the long
baselines associated with the scattering disc to be used for
localization \cite{Brisken2010, 2008MNRAS.388.1214W,
  2011MNRAS.416.2821D, Pen2012}. It is now possible to image the
interstellar scattering disk of pulsars at low frequencies, and it may
be possible to use these as billion km baselines to extract up to
nanoarcsecond astrometric information \cite{Pen2013}.

\section{Pulsars whose distances are poorly known}  
\label{PoorlyKnownDistances}

If the pulsar distance $r_{\alpha}$ is poorly known relative to
$\lambda_{gw}/(1+\hat{n} \cdot\hat{r}_{\alpha})$, then ${\cal
  P}_{\alpha}$ becomes a random phase containing essentially no
information; the ${\cal A}$ term is washed out, and only the ${\cal
  B}$ term remains in Eq.~(\ref{deltatDeriv}).  Such pulsars no longer
contribute diffraction-limited information, but all is not lost!

We start by sketching the key ideas, roughly. Consider, as a concrete
example, a GW of the form 
\begin{equation}
  h_{ij}(t,\vec{x})=h\;{\rm sin}[2\pi f_{0}(z+t)]
  \left(
    \begin{array}{ccc} 
      0 & 1 & 0 \\ 1 & 0  & 0 \\ 0 & 0 & 0
    \end{array}
  \right)
\end{equation}  
where $h$ is a positive real number. This is a GW with frequency
$f_{0}$ and ``cross'' polarization, traveling in the
$\hat{n}=-\hat{z}$ direction (which means that the GW source is
located at the north pole of the celestial sphere).  Then, using
Eqs.~(\ref{hij}, \ref{deltat(t)}, \ref{deltat(f)}) we find that 
the corresponding pattern of GW-induced timing residuals
is given by
\begin{equation}
  \label{deltat_alpha(t)}
  \delta t_{\alpha}(t)=-\frac{h\,(1+{\rm cos}\,\theta_{\alpha})
    {\rm sin}\,2\varphi_{\alpha}}{2\pi f_{0}}
  {\rm Re}\big\{[1-{\cal P}_{\alpha}(f_{0})]{\rm e}^{-2\pi i
    f_{0}t}\big\}.
\end{equation}
The first thing to notice is that this expression divides the
celestial sphere into 4 slices: $0<\varphi_{\alpha}<\frac{\pi}{2}$,
$\frac{\pi}{2}<\varphi_{\alpha} <\pi$,
$\pi<\varphi_{\alpha}<\frac{3\pi}{2}$ and
$\frac{3\pi}{2}<\varphi_{\alpha} <2\pi$. Thus, at time $t=0$ we see
that all of the pulsars in the slices
$0<\varphi_{\alpha}<\frac{\pi}{2}$ and
$\pi<\varphi_{\alpha}<\frac{3\pi}{2}$ exhibit {\it early} pulse
arrivals ($\delta t_{\alpha}<0$), whereas all of the pulsars in the
slices $\frac{\pi}{2}<\varphi_{\alpha}<\pi$ and
$\frac{3\pi}{2}<\varphi_{\alpha}<2\pi$ exhibit {\it late} pulse
arrivals ($\delta t_{\alpha}>0$); and at time $t=(2f_{0})^{-1}$ (half
a GW period later), the situation is reversed: all of the pulsars in
the slices $0<\varphi_{\alpha}<\pi/2$ and
$\pi<\varphi_{\alpha}<3\pi/2$ exhibit {\it late} pulse arrivals
($\delta t_{\alpha}>0$), whereas all of the pulsars in the slices
$\pi/2<\varphi_{\alpha}<\pi$ and $3\pi/2<\varphi_{\alpha}<2\pi$
exhibit {\it early} pulse arrivals ($\delta t_{\alpha}<0$). Note that
this is true {\it regardless} of the values of the unknown phases
${\cal P}_{\alpha}(f_{0})$ (contributed by the ``pulsar term'' in the
each pulsar's timing residual), since the real part of $[1-{\cal
  P}_{\alpha}(f_{0})]$ will always be positive (or, more correctly,
non-negative). The point is that the GW source lies at the point on
the celestial sphere where these four slices meet; so by localizing
the meeting point, we also localize the GW source. As we explain
below, among pulsars with poorly known distances, those close to the
GW source ($\theta_{\alpha}\ll1$) and far from the Earth
($r_{\alpha}\gg\lambda_{gw}/2\pi$) contribute particularly strongly to
the angular localization of the GW source.  To get an initial sense of
why this is true, first ignore the factor ${\rm Re}\{\ldots\}$ in Eq.
(\ref{deltat_alpha(t)}), and focus on the remaining angular dependence
$\kappa(\theta_{\alpha},\varphi_{\alpha})=(1+{\rm cos}\,
\theta_{\alpha}){\rm sin}\,2\varphi_{\alpha}$ \cite{hellings83}. Note
that $\kappa(\theta_{\alpha},\varphi_{\alpha})$ is singular at the GW
source's location ($\theta_{\alpha}=0$) because, although it does not
diverge as $\theta_{\alpha}\to0$, it does not vanish either, and
instead $\kappa$ approaches a different finite non-zero value
depending on the azimuthal direction $\varphi_{\alpha}$ along which we
take the limit $\theta_{\alpha}\to0$.  In other words,
$\kappa(\theta_{\alpha},\varphi_{\alpha})$ is singular in the sense
that, although $\kappa$ itself does not diverge, its azimuthal
($\varphi_{\alpha}$) derivative {\it does} diverge as
$\theta_{\alpha}\to0$.  By contrast, the full expression
(\ref{deltat_alpha(t)}) for $\delta t_{\alpha}$ is not singular:
including the factor ${\rm Re}\{\ldots\}$ smooths out the singularity
in $\kappa(\theta_{\alpha},\varphi_{\alpha})$ since, when
$\theta_{\alpha}$ is sufficiently small, $[1-{\cal P}_{\alpha}]$
vanishes as $\theta_{\alpha}^{2}$.  This is because, when the pulsar
is close enough to the GW source on the celestial sphere, its pulses
move along with the GW itself, with negligible change in their
relative phase (like a surfer riding an ocean wave); the ``earth
term'' and ``pulsar term'' then cancel, and no timing residual is
measurable.  If the pulsar is sufficiently far away from the Earth
($r_{\alpha}\gg\lambda_{gw}/2\pi$), this effect only kicks in for
small angular separations
$\theta_{\alpha}\lesssim\sqrt{\lambda_{gw}/(2\pi r_{\alpha})}$.  In
other words, when $\lambda_{gw}/(2\pi r_{\alpha})\ll1$,
Eq.~(\ref{deltat_alpha(t)}) is {\it quasi}-singular at
$\theta_{\alpha}=0$; it becomes {\it genuinely} singular in the limit
$\lambda_{gw}/(2\pi r_{\alpha})\to0$.  The fact that $\delta
t_{\alpha}$ varies rapidly (with ${\rm cos}\,2\varphi_{\alpha}$
dependence) around a tiny circle of radius $\sqrt{\lambda_{gw}/(2\pi
  r_{\alpha})} \lesssim\theta_{\alpha}\ll1$ surrounding the
quasi-singularity is the key to localizing the GW source.

To see this in more detail, split the pulsar directions
$\hat{r}_{\alpha}$ into components parallel and perpendicular to the
GW source direction $\hat{z}$:
$\hat{r}_{\alpha}=\hat{\rho}_{\alpha}{\rm sin}\,\theta_{\alpha}
\!+\!\hat{z}\,{\rm cos}\,\theta_{\alpha}$.  Now approach the
quasi-singularity in two steps: first consider the ``weaker'' limit
$\sqrt{\lambda_{gw}/(2\pi r_{\alpha})}\lesssim \theta_{\alpha}\ll1$ in
which $\theta_{\alpha}$ is small but $[1-{\cal P}_{\alpha}]$ is {\it
  not}; then proceed to the ``stronger'' limit $\theta_{\alpha}\ll
\sqrt{\lambda_{gw} /(2\pi r_{\alpha})}\ll1$ in which $\theta_{\alpha}$
and $[1-{\cal P}_{\alpha}]$ are {\it both} small.  In the weaker
limit, Eq.~(\ref{deltat(f)}) becomes
\begin{equation}
  \label{deltat_approx}
  \delta\tilde{t}_{\alpha}(f)\approx\frac{i}{\pi f}
  \tilde{h}_{ij}(f)\hat{\rho}_{\alpha}^{i}\hat{\rho}_{\alpha}^{j}
  [1-{\cal P}_{\alpha}(f)]
\end{equation}
while Eq.~(\ref{deltatDeriv}) becomes
\begin{equation}
  \label{deltatDeriv_approx}
  \frac{\partial[\delta\tilde{t}_{\alpha}(f)]}{\delta\gamma^{\,\bar{\mu}}}
  =\frac{2i}{\pi f}\frac{C_{\alpha\bar{\mu}}(f)}
  {\theta_{\alpha}}[1-{\cal P}_{\alpha}(f)]
\end{equation}
where 
\begin{equation}
  C_{\alpha\bar{\mu}}(f)\equiv\tilde{h}_{ij}(f)
  \hat{\rho}_{\alpha}^{i}\big[\epsilon^{j}_{\;kl}
  \hat{n}^{k}\hat{m}_{\bar{\mu}}^{l}\!+\!
  \hat{\rho}_{\alpha}^{j}\hat{\rho}_{\alpha}^{k}
  \hat{n}^{l}\hat{m}_{\bar{\mu}}^{m}\epsilon_{klm}\big].
\end{equation}
Thus (still in the weaker limit) we have:
\begin{equation}
  \label{quasi_singularity_limited}
  \frac{\Gamma_{\bar{\mu}\bar{\nu}}^{\alpha}}{{\rm SNR}_{\alpha}^{2}}
  \approx\frac{4}{\theta_{\alpha}^{2}}\,
    \frac{C_{\alpha\bar{\mu}}(f_{gw})C_{\alpha\bar{\nu}}(f_{gw})^{\ast}}
    {\big|\tilde{h}_{ij}(f_{gw})\hat{\rho}_{\alpha}^{i}
        \hat{\rho}_{\alpha}^{j}\big|^{2}}.
\end{equation}
Since the second fraction on the right-hand side of this expression is
generically ${\cal O}(1)$, this says that when a pulsar is near (but
not {\it too} near) a GW source on the sky, its contribution to
$\Gamma_{\bar{\mu}\bar{\nu}}$ is typically
$\Gamma_{\bar{\mu}\bar{\nu}}^{\alpha} \sim(4/\theta_{\alpha}^{2}){\rm
  SNR}_{\alpha}^{2}$.

In the stronger limit $\theta_{\alpha}\ll\sqrt{\lambda_{gw}/ (2\pi
  r_{\alpha})}\ll1$, Eqs. (\ref{deltat(f)}) and (\ref{deltatDeriv})
imply that $\delta\tilde{t}_{\alpha}(f)$ and
$\partial[\delta\tilde{t}_{\alpha}(f)]/\partial\gamma^{\,\bar{\mu}}$
are smooth and vanishing at $\theta_{\alpha}=0$.  So as
$\theta_{\alpha}$ decreases, $\Gamma_{\bar{\mu}\bar{\nu}}^{\alpha}$
initially increases as $1/\theta_{\alpha}^{2}$, and then drops to
zero; in between it attains a maximum value:
\begin{equation}
  \label{Gamma_max}
  \Gamma_{\bar{\mu}\bar{\nu}}^{\alpha}\sim\frac{8\pi r_{\alpha}}
  {\lambda_{gw}}{\rm SNR}_{\alpha}^{2}
\end{equation}
at a separation angle $\theta_{\alpha}\approx \sqrt{\lambda_{gw}/(2\pi
  r_{\alpha})}\ll1$.

\begin{figure}
  \begin{center}
    \includegraphics[width=3.1in]{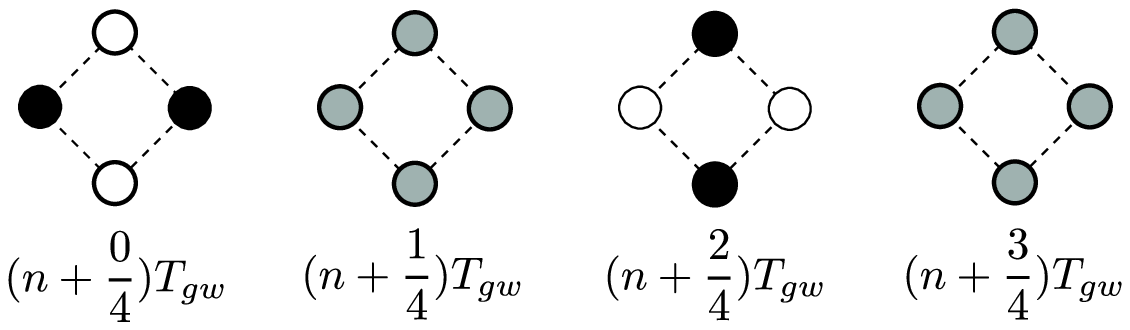}
  \end{center}
  \caption{The 4 circles represent 4 pulsars that form a small square
    on the sky.  The timing residuals of all 4 pulsars are oscillating
    with the same amplitude and period $T_{gw}$, but different phases;
    each pulsar's oscillation is $180^{o}$ out of phase with its
    2 nearest neighbors.  (This figure depicts this by showing 4
    different moments in the oscillation cycle: black, white, or grey
    circles indicate that, at that moment, the pulses are arriving
    early, late, or ``on time,'' respectively.)  This signature
    indicates that the square contains a GW point source.}
\label{fig1}
\end{figure}

Now consider multiple GW sources.  We can repeat the previous
section's counting argument, except that we must now include the $N$
unknown pulsar distances when we are counting the parameters needed to
specify the pattern of timing residuals; with this modification, we
find that a PTA which monitors $F$ different GW frequency bins can
completely characterize up to an average of $(2N/7)(1-1/2F)$
``chirping'' sources [or $(2N/6)(1-1/2F)$ monochromatic sources] per
bin.  Note that, although we didn't know the pulsar distances {\it a
  priori}, they are determined, in principle, by the fit \footnote{In
  the special case that the gravitational wave signal is due to a
  single monochromatic gravitational plane wave, this fit is highly
  degenerate, since each $r_{\alpha}$ is only determined modulo
  $\lambda_{gw}/ (1+\hat{n}\cdot\hat{r}_{\alpha})$; but, if the GWs
  are significantly non-monochromatic ({\it e.g.}  ``chirping''),
  significantly non-planar \cite{DengFinn} or, most importantly, if
  there are two or more GW signals propagating in different
  directions, this degeneracy is broken.}  \footnote{After completion
  of this work, we learned that Ref.~\cite{CorbinCornish} had recently
  emphasized a similar point.}.

If the PTA can disentangle and characterize the individual sources,
how well can they be angularly localized?  To answer this question,
one should ask, for each combination of pulsar and GW source, whether
the fit to the timing residuals has determined $r_{\alpha}$ accurately
or poorly relative to $\lambda_{gw}/(1+\hat{n}\cdot\hat{r}_{\alpha})$;
roughly speaking, if $r_{\alpha}$ has been determined accurately then
we expect the pulsar will contribute diffraction limited angular
information as described by Eq.~(\ref{diffraction_limited}); and if
$r_{\alpha}$ has been determined poorly then we expect the pulsar will
contribute ``quasi-singularity limited'' angular information
$\Gamma_{\bar{\mu}\bar{\nu}}^{\alpha}$ for that source, as described
by Eqs.~(\ref{quasi_singularity_limited}) and (\ref{Gamma_max}).
Consider the localization of a GW source when all of the pulsar's have
poorly known distances; as explained above, the quasi-singular pattern
of timing residuals implies that the angular localization will be
dominated by the pulsars that are close to that source on the
celestial sphere; in particular, it is roughly set by the smallest
quadrilateral of pulsars that encircles the source on the celestial
sphere, down to a limiting angular resolution of roughly
$\delta\theta\sim(1/{\rm SNR}_{\alpha})
\sqrt{\lambda_{gw}/d_{pulsar}}$ [the more precise statement is given
by Eqs.~(\ref{quasi_singularity_limited}) and (\ref{Gamma_max})].  To
understand this behavior, consider the example in Fig.~\ref{fig1}.

\section{Discussion} 
\label{Discussion} 

As we have seen, if a PTA is confronted with too many gravitational
wave GW sources, it will become ``confused'' (unable to disentangle
the sources).  It was previously assumed ({\it e.g.}\ in \cite{SVV}),
as a rule of thumb, that PTAs would become confused when there was
more than one GW source per frequency bin.  In this paper, we have
derived the confusion limit, and shown that the actual result is
different: a PTA containing N pulsars does not become confused until
there are 2N/7 sources per frequency bin (or slightly less, if the
pulsar distances are not accurately known); until this threshold is
crossed, the PTA is sensitivity limited, not confusion limited.  In
other words, a PTA with many pulsars is much less confused than the
naive rule of thumb would suggest.  To translate this into concrete
terms, see the lower left hand panel of Figure 2 in Ref.~\cite{SVV}.
This figure estimates the expected number of GW sources above a
certain minimum timing residual threshold.  From it, we see that: the
expected number of GW sources per frequency bin with timing residuals
above 100 ns (a relatively near-term/achievable threshold) is well
below 0.1 in every bin; the expected number of GW sources per
frequency bin with timing residuals above 10 ns (a rather ambitious
threshold) is well below 10 in every bin; and even the expected number
of GW sources per frequency bin with residuals above 1 ns (a very
ambitious/futuristic threshold) is well below 100 in every bin.  Thus,
even in the rather ambitious/futuristic case in which the PTA can
detect GW sources with timing residuals as small as 10 ns, if the PTA
contains $\gtrsim 35$ quiet pulsars, it will be sensitivity limited
rather than confusion limited, even in its lowest frequency bins.  And
even in the {\it very} ambitious/futuristic case in which the PTA can
detect GW sources with timing residuals as small as 1 ns, if the PTA
contains $\gtrsim 350$ quiet pulsars (see e.g. \cite{IPTA}), it will
be sensitivity limited rather than confusion limited, even in its
lowest frequency bins.  (Let us put these numbers in some context.  At
the moment, there are about 20 pulsars with rms timing residuals less
than 1 $\mu$s, and a few pulsars with rms timing residuals as small as
100 ns \cite{Demorest09}.  In the nearer term, the Parkes Pulsar
Timing Array is aiming to monitor an array of 20 pulsars, each with
rms residuals better than 100 ns over a 5 year timescale
\cite{Manchester2008}.  And in the future, the Square Kilometer Array
project could detect more than 20,000 pulsars, including hundreds of
pulsars with rms timing residuals that match or surpass the best few
pulsars currently available \cite{Kramer, IPTA, SesanaVecchio}.)
Thus, it is a very plausible possibility that future PTAs will be
sensitivity limited, rather than confusion limited, even at their
lowest frequencies.  In this case, the traditional picture of a PTA as
a stochastic background detector will be incorrect, and there will be
a significant advantage to analyzing the data via matched filtering
(as is done in other gravitational wave experiments such as LIGO
\cite{LIGO}) and thinking of the PTA as a point source telescope,
rather than analyzing it as if it were detecting a stochastic
background.

In the previous sections, we have attempted to clarify the limits on
the capabilities of PTAs and, in particular, how these limits depend
on factors such as the SNR distribution of the GW sources, the number
and angular distribution of the pulsars relative to the GW sources,
the distances to the pulsars and the precisions of those distances.
Our results for the angular resolution were obtained by Fisher matrix
methods, assuming gaussian, stationary noise.  For this reason, we
must remember that in general these results must be interpreted as
{\it bounds} on (rather than {\it estimates} of) the angular
resolution.  For high-SNR sources, these bounds will be saturated, and
may be reinterpreted as actual estimates; for moderate-SNR sources
they will still provide useful quantitative guidelines; but for
low-SNR sources, it is especially important to remember that our
formulae represent bounds rather than estimates, since the Fisher
bounds will become increasingly ``loose'' ({\it i.e.}\ non-saturated)
in the low SNR regime \cite{Lee}.  (Note that the relevant SNR here is
the {\it total} SNR of the source in the PTA, which can be high even
if the SNR per pulsar is not.)


In addition to the analytical formulae presented and interpreted
earlier, it is worth mentioning one other rule of thumb to keep in
mind when estimating the angular resolution of a PTA.  Generically,
since most pulsars tend to be located in the galactic plane, one
expects the spatial resolution to be best for GW sources near the
galactic pole if pulsar distances are accurately known, and best for
GW sources in the galactic plane if pulsar distances are poorly known.

This paper has focused on the angular resolution of a PTA in the
regime where the GW sources are not confused; and on the conditions
that need to be satisfied in order to be in this regime.  In the
opposite (fully confused) regime, the gravitational wave signal may be
treated as a Gaussian random field; this situation has been
extensively studied in the literature.  The intermediate regime in
between these two limits is also very interesting, but clearly more
complicated than either limit, and will be left for future work.

Other interesting problems for future work include: (i) ``tightening'' the
angular resolution limits at low SNR, beyond the limits obtained by
Fisher matrix methods \cite{Lee}; (ii) extending this work to GW point
sources that are near enough that their wavefront curvature is
significant \cite{DengFinn}; (iii) determining the circumstances in
which pulsar distance determination by GW fitting can compete with
more traditional methods ({\it i.e.}\ VLBI, cyclic spectroscopy or
timing parallax), see \cite{CorbinCornish}; (iv) clarifying the
statistics of GW sources which are anomalously well characterized
because they are fortuitously located relative to one or several
pulsars on the sky; (v) quantifying the gain from matched filtering
(with quasi-singular filters in particular) compared to traditional
stochastic correlation analysis, even when the PTA {\it appears}
source confused; (vi) understanding the predicted distribution of 
frequency derivative $\dot{f}$ among the GW sources relevant
to PTAs, and the implications of this distribution for PTA 
GW telescopes.

{\bf Acknowledgements.}  We are grateful to Chris Hirata, Neil Turok,
and Michael Kramer for valuable conversations.  LB acknowledges
support from the CIFAR JFA.  UP received NSERC support for this work.

\end{document}